\documentstyle{article}
\def \D {\Delta}

\def \E {{\bf E }}

\def \l {\lambda}
\def \s {\sigma}
\def \T {{\hbox { Tr }}}
\def \d {{\hbox {d}}}

\def \a {\alpha}

\def \de {\delta}
\def \W {{\cal W}}
\def \V {{\cal V}}

\begin{document}

\begin{large}
\centerline{ON ASYMPTOTIC SOLVABILITY }
\centerline{OF RANDOM GRAPH'S LAPLACIANS}

\vskip 0.5cm
\centerline{A. Khorunzhy$^{\star,\#}$ and V. Vengerovsky$^{\star}$}

\vskip 0.3cm
\centerline{$^{\star}$Institute for Low Temperature Physics,
Kharkov, UKRAINE}
\centerline{$^{\#}$Facult\'e de Math\'ematiques,
Universit\'e Paris-VII,  FRANCE}
\vskip 0.5cm

\vskip 0.5cm
\end{large}
{\bf Abstract} We observe that the Laplacian
of a random graph $G_N$ on $N$
vertices represents an explicitly solvable
model
in the limit of infinitely increasing $N$.
Namely, we derive recurrent relations for the
limiting averaged moments
of the adjacency matrix of $G_N$ when $N\to\infty$.
These relations allow one to study  the
corresponding eigenvalue distribution function; we show that
its density has a infinite support in contrast to the case of the ordinary
discrete Laplacian.

\begin{large}
\vskip 0.5cm
\noindent  Key words:
\noindent {\it graph Laplacian, random graphs,
random matrices, eigenvalue distribution,
walks on trees}

\vskip 0.5cm
The spectral theory of graphs
attracts more and more attention in mathematical physics
(see e.g. \cite{CdV,N}).
It is widely believed that the spectral properties
of graphs are related with their geometry that is an intriguing
assertion from
various points of view.
However
in general, without the knowledge of the graph structure,
one can prove just  abstract theorems with not large number
of concrete results.
In this connection, the study of the spectrum
of random graphs, as the typical in certain sense objects, can be useful.
We confirm this assumption by showing
that the models related with large random graphs are explicitly solvable.

Given a set $\V_N$ of $N$ vertices $v_1, \dots v_N$, one can assume that
some pairs of them are either joined by one
undirected edge or are not joined.
Then $\V_N$ and the set $E_N$ of existing edges determines the graph
$\Gamma_N=(\V_N,E_N)$.
In this case the graph Laplacian $\D_N$ acting on functions
$f(v), v\in \V_N$
is determined by the formula
$$
\D_N f (v_i) = \sum_{j: v_j\sim v_i} [f(v_i)-f(v_j)],
\eqno (1)
$$
where the sum runs over the vertices joined with $v_i$.

It is clear that $\D_N$ has $N\times N$ real symmetric matrix
and
$$
\D_N = V^{(N)} - A^{(N)},
$$
where $A^{(N)}$ is the adjacency matrix of the graph $\Gamma_N$
$$
A^{(N)}_{ij} = \cases{1, & if $v_i\sim v_j$,\cr
0, & otherwise, \cr}
\eqno (2)
$$
and $V^{(N)}$ is the diagonal matrix such that
$V^{(N)}_{ii} = \deg (v_i)$ is the number of edges attached to $v_i$.
Note that $A^{(N)}_{ii}=0$ and that
$$
\D_N = P_N [A^{(N)}]^2 P_N - A^{(N)},
\eqno (3)
$$
where $P_N$ is the orthogonal projection onto the diagonal.
Often the set of eigenvalues $\l_1\le\dots\le \l_N$
of $A_N$ is called the graph's spectrum \cite{CDS}.
Firstly we  concentrate on this problem and turn to
the spectrum of $\D_N$ at the end of this article.

Our principal goal is to describe results
obtained for  the adjacency  matrix
$A^{(N,p)}$ of a random graph $G_{N}^{(p)} = (\V_N,E_{N}^{(p)}(\omega))$.
We are related with the well-known model (see e.g. \cite{B}),
where  $\{G_{N}^{(p)}\}$
represents the ensemble of graphs
where each edge, independently
from the others, is either present
or absent
with probabilities $p/N$, $0<p\le N$
and $1-p/N$, respectively.
Thus we get a real symmetric matrix $A^{(N,p)}$
with the entries $\{A^{(N,p)}_{ij}, i\le j\}$ determined
as a family of jointly independent
random variables such that
$$
 A^{(N,p)}_{ii} = 0 {\hbox{ and }}
A^{(N,p)}_{ij} = \cases{1, & with probability $p/N$,\cr
0, & with probability $1-p/N$. \cr}
\eqno (4)
$$
We are interested in asymptotic behaviour in the limit $N\to\infty$
of the moments
$$
M^{(N,p)}_s = \E \{ N^{-1} \T A^s \}, \quad A\equiv A^{(N,p)},
\eqno (5a)
$$
where
$$
N^{-1} \T [A^{(N,p)}]^s = {1\over N} \sum_{x_i=1}^N
A_{x_1x_2} A_{x_2x_3} \cdots A_{x_{s-1}x_1}.
\eqno (5b)
$$
We consider for simplicity the case of $p=1$. Our main result is that
$$
\lim_{N\to\infty}M^{(N,1)}_s = \cases{ m_{k}, & if $s=2k$\cr
0, & if $s=2k+1$\cr}
$$
and $m_k$ can be found  from equality
$$
m_k = \sum_{r=0}^k W_k(r)
\eqno (6a)
$$
where the numbers $W_k(r)$, $k\ge r \ge 0$ are given by the following
recurrent relations
$$
W_u(v) =
 \sum_{i=1}^v
\sum_{j=v-i}^{u-i}
\sum_{l=0}^{u-i-j}
W_{u-i-j}(l)
{l+i-1\choose i-1}
{v-1\choose i-1}
W_j(v-i),
\eqno (6b)
$$
supplied
with the initial condition $ W_j(0)=\de_{j0} $.

Derivation of (6)
essentially uses the fact that $A^{(N,p)}$ (4) is the matrix whose
entries are jointly independent (excepting the symmetry condition)
random variables. For such class of matrices, the limits of
moments (5) were studied first by E. Wigner \cite{W}.
The ensemble considered in \cite{W} is somewhat different from (4),
and the method proposed there and
then developed
in \cite{FK}  is applicable to (4) in the asymptotic regime $p\gg 1$
only.

The principal observation of \cite{W} and \cite{FK} is that
the sum over
the sequences $X_{2k} \equiv (x_1,x_2, \dots, x_{2k-1},x_1)$ in (5a)
can be replaced by the sum over the
equivalence classes of the walks $X_{2k}$;
these classes are labelled by the plane rooted trees $\tau\in T_k$
drawn in the upper half-plane with the help of $k$ edges.

In paper \cite{K1} the method of \cite{W} and \cite{FK}
was further improved up to the stage when $p\sim 1$
can be also included into consideration.
Summimg up the arguments  of \cite{W,FK} and \cite{K1},
we claim that
$$
m_k = \sum_{q=0}^{k-1} \sum_{\tau \in T_{k-q}}
U(k;\tau),
\eqno (7)
$$
where
$U(k;\tau)$ is the number of all possible
different walks of $2k$ steps such that cover the tree $\tau\in T_{k-q}$
according to the
following

{\bf Riding rule:} {\it
the walk starts and ends at the root and  each edge
of the tree is passed even positive number of times;
when
there is
a choice where to move further,
the walk chooses either one of the edges that has been
already passed or the most left edge
among those that have not yet been  passed.}

Let us call the walks of this type the {\underline {even ordered walks}
covering the tree}.

{\it Demonstration of (6)}.

It is convenient to imagine
that some particle moves in the upper half-plane.
It starts at the root
$\rho$, then makes $2k$ steps and returns to the root;
each step either produces a new vertex or
ends at the vertex that already exists
and is connected by the present vertex by an edge.
The condition
is that each step made in one direction
is repeated in inverse direction.

It is easy to see that
there exists a tree $\tau \in T_{k-q}$ (and the only one)
that the walk
of the particle represents one of the coverings of $\tau$
satisfying the riding rule.

Let us introduce the set
$
\W_u(v)
$
of all possible walks of $2u$ steps
that  return to the root
$v$ times and denote by $W_u(v) = \vert \W_u(v)\vert $
the number of such walks. With this definition in mind,
it is not hard to derive (6) from (7).

Indeed, regarding the set of walks $\W_u(v)$, let us denote
their first step by $(\rho,\a)$  and consider those walks that
pass this edge $2i\ge 2$ times.
It is easy to see that $\a$ by itself can be regarded
as the root of a subwalk that starts and ends at $\a$
and never visits $\rho$.
Chronologically, this trajectory consists of possibly several pieces.
Let us call their union the first subwalk.
Similarly, the second subwalk starts and ends at $\rho$
and never visits $\a$.

Let us assume that
the second subwalk belongs to $\W_j(v-i)$ and the
first subwalk belongs to $\W_{u-i-j}(l)$.
Then (6b) follows, where  in the product, the second factor
represents the number of possibilities to
perform $i-1$ steps from $\a$ to $\rho$
under the condition that we arrive $l$ times to $\a$
by the first subwalk; the last passage $\a\to\rho$
is made after that the first subwalk is completed.
The third factor in (6b)
corresponds to $i-1$ steps $\rho\to\a$
made after $v-i$ arrivals to $\rho$ by the second subwalk.$\Box$


In the following
remarks we
 briefly discuss the result obtained

1. The first conclusion  relates the maximal eigenvalue
of $A^{(N,1)}$. It is clear that
the number of non-zero elements in a row is
asymptotically Poissonian when $N\to\infty$.
Thus, with non-zero probability, in $(\V_N, E_{N}^{(1)})$ there exists
a vertex that has large number of edges attached to it.
This explains why $\Vert A^{(N,1)}\Vert \to\infty$
(see also \cite{K1}) while the spectrum of the ordinary
discrete Laplacians
(or more generally,
of the regular graphs) remains bounded.

The stress of the present article is that
there is an infinite number of the eigenvalues
of $A^{(N,p)}$
that tend to infinity as $N\to\infty$.
We derive this conclusion from
relations (6) and the representation
$$
M^{(N,p)}_s= \E \int \l^s \d \s_{N,p}(\l),
\eqno (8)
$$
where
$\s_{N,p}(\l) = \s(\l, A^{(N,p)})$
is the normalized eigenvalue counting
function
$ \s(\l) = \# \{\l_j \le \l\} N^{-1} $.

Basing on (6b),
one can easily show that $W_k(r)\le (C_1k)^{2r}$
where $C_1$ is some constant.
Then we derive from (6a)  the estimate
$$
m_{2k}\le (C_2 k)^{2k}
$$
that implies the uniqueness of the limiting measure
$\s^{(1)}= \lim_{N} \s_{N,1}$.

>From another hand, it is not hard to derive the estimate
$$
m_{2k}\ge (k/2)^{k/2}(1+o(1)), k\to\infty.
\eqno (9)
$$
Indeed, let us take the set  $\W_k(k)$
and consider the walks that cover one
particular tree $\tau'_{k/2}$ of $k/2$ edges,
namely that has all $k/2$
edges attached to the root (here we assume $k=2l$ for simplicity).
Among these walks there are those that
pass each edge four times. It clear that the number of
these walks is $(k/2)!$ that results in (9).

Inequality (9) implies that the measure $\d\s^{(1)}$ has
an infinite support.
This means that the mathematical expectation of the number
of eigenvalues
that go to infinity is of the order $O(N)$ as $N\to\infty$.

2. Let us note that one can easily apply our method to
study the limits $m_k^{(p)}$ of  moments
(8) with arbitrary constant number $p$.
In paper \cite{FM} the arguments have been presented
showing that there exists a critical value $p_c \sim 2$
that separates two different pictures of the
limiting spectra of $A^{(N,p)}$. This is supported
by the facts from the random graphs theory stating that
$p_c'=1$ is critical for existing of an infinite
connected component in $(\V_N, E_{N}^{(p)})$ as $N\to\infty$
(see e.g. \cite{B}).
Our preliminary numerical studies of the moments
$m_k^{(p)}$ show that
$\s^{(p)}$ is not sensitive up to the level
that  reflects the role of
$p_c$ or $p_c'$.

3. It should be mentioned that the moments (7) we compute
represent a special kind of the matrix integrals.
If one considers the matrices $A_N$ (4) as having the Gaussian
invariant distribution,
then the corresponding moments  (5) can be computed explicitly
for fixed $N$
and the formulas obtained are related with the
number of maps drawn on the surfaces
of given genus (see e.g. \cite{Z} for an accessible introduction).

In this aspect, our model characterises another class of
random matrices with independent entries that have
explicitly countable integrals over them
(see \cite{SS} for related recent
results concerning their asymptotic properties).

4. Finally, let us indicate the way to find explicit relations for
the moments of the
random graph's Laplacian. Regarding definition (3),
we see that one faces the problem of computing the mixed moments
$$
\E\left\{A^{r_1}(x_1,x_2) [A^2(x_2,x_2)]^{s_1} \cdots
A^{r_j}(x_j,x_1) [A^2(x_1,x_1)]^{s_j}\right\}.
$$
One can apply again the method of \cite{W,FK} with the modification
that the trees are constructed with the help of edges
of two different colours and the edges of one of  the class
can be used as the leaves only.
Certainly, in the case of finite $p$ the computations
become more difficult than those of (6),
but the general approach of \cite{K1}
and the present article remains valid.

Note that the case of multiple edges can be also
included into consideration by allowing
$A_{ij}$ (4)  to take positive integer values. Moreover,
our method still works for
the random graphs with independent random weigths on edges;
then the adjacency matrix (2) takes the form
$B_{ij} = \Xi_{ij} A_{ij}$, where $\Xi_{ij}, i\le j $
are jointly independent random variables.
All these questions will be considered in more details
in separate publication.


\end{large}
\end{document}